\documentclass[10pt]{article}
\usepackage{url}
\usepackage{hyperref}
\hypersetup{
    colorlinks=true,
    linkcolor=blue,
    filecolor=magenta,      
    urlcolor=blue,
    pdftitle={Overleaf Example},
    pdfpagemode=FullScreen,
    }
\usepackage{latexsym}
\usepackage{amsmath, amsthm, amsfonts}
\usepackage{algorithm, algorithmic}  
\usepackage{graphicx}
\usepackage{listings}
\usepackage{blindtext}
\usepackage[a4paper, total={6.5in, 9.7in}]{geometry}
\usepackage{graphicx}
\usepackage[font=small]{caption}
\usepackage[utf8]{inputenc}
\usepackage[T1]{fontenc}
\usepackage{subfig}
\graphicspath{{Images/}}
\usepackage{tikz} 
\usepackage{siunitx}
\usepackage{pgfplots}
\pgfplotsset{compat=newest} 
\usepgfplotslibrary{units}
\title{\textbf{{\huge Setting Out a Software Stack Capable of \\ Hosting a Virtual ROS-based Competition}}}

\author{
  Nishesh Singh \\
  Software Developer, Mars Society South Asia (MSSA) \\
  {\tt nishesh.singh@learner.manipal.edu} \\}

\begin{document}
\maketitle
\begin{abstract}
Traditional academic competitions  that foster collaboration from student communities all around the globe have either been postponed indefinitely or canceled due to restrictions posed upon in-person gathering. Owing to this, virtual competitions are gaining importance as they provide the student community with an outlet for exposure and academic growth whilst having the convenience of being online and safe. Mars Society South Asia (MSSA) has developed a software stack capable of replicating on-site competition tasks via ROS-based simulations. The developed software \footnote{{\url{https://github.com/singhnishesh/Virtual-Mars-Rover-Challenge-2021}}.} enables users to perform necessary simulations without them having to interact with ROS or Gazebo. Currently, the software has been set out to be used for the Virtual Mars Rover Challenge (VMRC), a competition that focuses on simulating the next generation of Mars rovers and thereby providing the student community with a substitute to their on-site competitions. The objective of this white paper is to explain the software's architecture, installation, usage, maintenance, and ease of use which is the singular most important factor in software adoption.
\end{abstract}

{\bf Keywords:} ROS, Whitepaper, OpenAI Gym , \texttt{gym-gazebo}.

\section{Introduction}

Reinforcement Learning is a focused area of machine learning where an agent interacts with its environment and learns to take a series of actions to maximize the notion of cumulative reward.
To build upon recent progress in Reinforcement Learning, the research community needs good benchmarks on which to compare cutting-edge algorithms. OpenAI  \cite{brockman2016openai} aims to combine the best elements of these previous benchmark collections, in a software package that is maximally convenient and accessible. It is a standardized toolkit for comparing different Reinforcement Learning paradigms. The \texttt{gym} library is a collection of community-maintained test problems (environments) that have a shared interface allowing the user to test different algorithms. Unfortunately, \texttt{gym} does not facilitate the provision of interfacing with the Robot Operationg System (ROS) \cite{quigley2009ros}. Robot Operating System ROS is an open-source robotics middleware suite. Although ROS is not an operating system but a collection of software frameworks for robot software development, it provides services designed for a heterogeneous computer cluster such as hardware abstraction, low-level device control, implementation of commonly used functionality, message-passing between processes, and package management. Gazebo \cite{koenig2004design} is the default ROS simulator, while ROS serves as the interface for the robot. \texttt{gym-gazebo} \cite{zamora2016extending} is a complex piece of software that essentially acts as the OpenAI extension for Gazebo based environments. The extension enables us to use \texttt{gym} functionalities on Gazebo-based environments which simplifies the simulation process considerably.

\section{Background}

There are two basic entities: the environment which is the world the robot is present in and the agent which is the robot itself. The Agent-Environment Interaction (AEI) model \cite{deloach2006agent} is composed of three main elements: the Capability Model, the Environment Model, and a set of interactions between capabilities and environment objects. Essentially, agents possess capabilities that sense and act upon objects in the environment via interactions. The figure provided below depicts the integration of these three parts into the AEI Model. The top part of the figure represents the Capability Model, which defines capabilities as consisting of a set of actions, each of which has a single operation that interacts with environment objects. The bottom part of the figure captures the Environment Model, which includes an explicit environment that contains a set of environment objects (that includes agents) and their relationships. The environment objects are governed by processes that implement specific environmental principles. Interactions are defined by the intended effect of operations on environment objects.

\begin{figure*}[htbp]
\centering
\includegraphics[width=12cm]{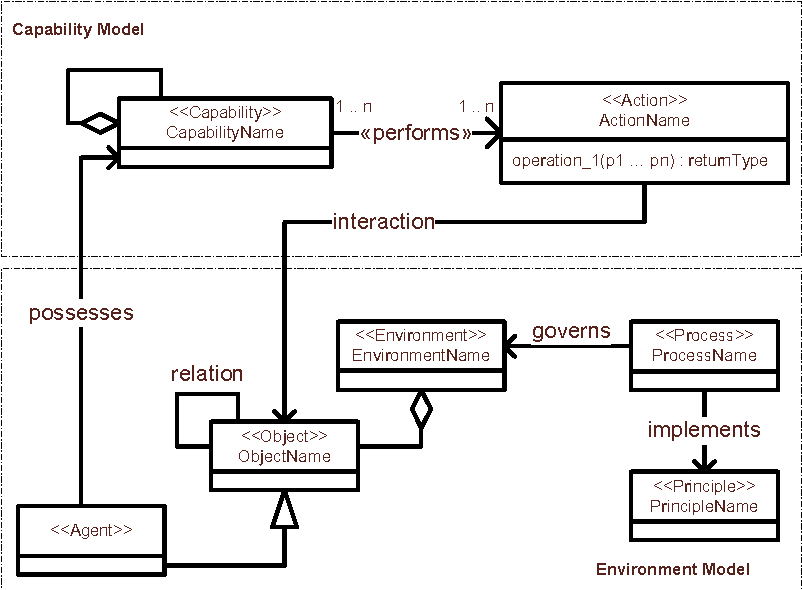}
\caption{Agent-Environment Interaction (AEI) model \cite{deloach2006agent}.}
\label{fig:AEI model}
\end{figure*} 
Env is the core \texttt{gym} interface. To interact with the environment via the agent, \texttt{gym} provides 3 main API methods that users of this class need to know are:
\begin{itemize}
\item \texttt{step(self, action)}: steps through the environment by one timestep and returns a \emph{done} flag after completion.
\item \texttt{reset(self)}: resets the state of the environment.
\item \texttt{render(self)}: renders one frame of the agent-environment interaction.
\end{itemize} \par
\texttt{gym-gazebo} provides the same functionalities for a ROS-based environment with the addition of a \texttt{get\_observation(self)} method that allows arbitrary repeated access to the current observation. The \texttt{gym.make(self,\emph{"environment\_name"})} method at its core calls the constructor corresponding to the registered environment .The pseudo-code below demonstrates how to iterate the agent through the environment for successive time steps. Furthermore, the code resets the environment to its initial state when a condition is satisfied.
\begin{lstlisting}[language=python]
import gym
env = gym.make('environment_name-v0')
while not done:
    action = env.action_space.sample()
    env.step(action)
    if(env.get_observation() > threshold):
        done = True
        observation = env.reset()
\end{lstlisting}
The developed software stack is centered around the \texttt{gym-gazebo} package and its functionalities. It incorporates some new modifications from our end while simplifying the installation process greatly. Furthermore, the modifications currently strictly adhere to the class of mobile ground robots and more specifically differential drive robots. Additions also include the inclusion of the setup file, a bash script that interacts with the user and upon successful execution automatically integrates a controller for differential drive wheel systems \footnote{\texttt{\url{{https://github.com/ros-controls/ros_controllers.git}}}}, gazebo plugins from Team Hector\footnote{\texttt{\url{https://github.com/tu-darmstadt-ros-pkg/hector_gazebo}}} which  currently contains a 6 wheeled differential drive plugin, an IMU sensor plugin, an earth magnetic field sensor plugin, a GPS sensor plugin and a sonar ranger plugin, a generic keyboard teleoperation package for twist robots \footnote{\texttt{\url{https://github.com/ros-teleop/teleop_twist_keyboard.git}}} , and other necessary sensor plugins with the preliminary ROS package of the robot provided by the software operator. \par The user input to the software includes the robot link's physical properties such as geometry, color, mass, and inertia, and the collision properties of the link which is encapsulated in the Universal Robot Description Format (URDF) file. To generate the URDF the user is encouraged to use the Solidworks to URDF exporter \footnote{\texttt{\url{http://wiki.ros.org/sw\_urdf\_exporter}}}, a SolidWorks add-in that allows for the convenient export of SolidWorks parts and assemblies into a URDF file. The exporter will create a ROS-like package that contains a directory for meshes, textures and robots (URDF files). For single SolidWorks parts, the part exporter will pull the material properties and create a single link in the URDF. For assemblies, the exporter will build the links and create a tree based on the SolidWorks assembly hierarchy. The exporter can automatically determine the proper joint type, joint transforms, and axesex
\section{Architecture}
The architecture consists of 3 independent software packages: OpenAI Gym, ROS and Gazebo.  Environments developed in OpenAI Gym interact with ROS, which is the connection between \texttt{gym} itself and the Gazebo simulator.
\begin{figure}[htbp]
\centering
\includegraphics[width=9cm]{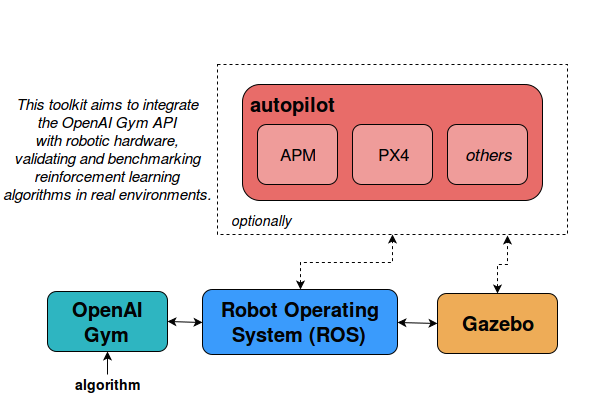}
\caption{Simplified software architecture and interaction \cite{zamora2016extending} }
\label{fig:Architecture}
\end{figure}
The training script uses the \texttt{get\_observation(self)} method that allows arbitrary repeated access to the current observation perceived by the agent and publishes this data using the \texttt{step(self, action)} method. This script is the primary control code and users need to edit this code in a manner that fits their requirements and functionalities. 

\subsection{Installation}
We have deployed an automatic installation script known as the helper script to simplify the setup process of the complex stack by a great extent. This bash script upon execution builds the software from source post the sequential installation of all required dependencies, Linux / Python3 packages, platform-dependent / independent tools and available catkin workspace(s). This leads
to an easier and faster installation process , which  facilitates providing assistance to the non Linux / ROS experienced part of the community.
\subsection{Setup}
The setup file is a bash script that interacts with the user and upon successful execution automatically integrates the aforementioned plugins with the preliminary ROS package of the rover provided by the software operator. This setup file is primarily designed for competition tasks that require a teleoperation task from the point of view of the agent. The keyboard teleoperation package  takes input from the keyboard and constantly publishes it to the \texttt{geometry\_msgs/Twist} message type provided by the differential drive controller. The input to the ROS topic \texttt{/cmd\_vel} is both the linear and angular component of the velocity via the  \texttt{geometry\_msgs/Twist} message type.
\subsection{Graphical User Interface}
Owing to the restrictions posed by Gazebo like the inability to view the task environment from the point of view of the agent using on-board cameras present, teleoperation tasks require an additional toolkit for 3D visualization. Luckily, ROS has such a toolkit namely RVIZ. RVIZ is a ROS graphical interface that allows the user to visualize a lot of information using plugins for many kinds of available topics making it suitable for our purpose.
\subsection{Command Line Interface (CLI) commands}
Most of the times, ending or killing the simulation does not shut out \texttt{rosmaster} or \texttt{gzserver} and these processes keep rendering output thereby slowing computation and causing the system to crash. \texttt{gym-gazebo} installation creates an alias namely \texttt{killgazebogym}. This CLI command makes sure to kill all existing background processes before the user starts new tests.

\section{Environments}
The stack currently provides the following two environments:
\begin{itemize}
\item Leo Rover \footnote{\texttt{(\url{https://github.com/LeoRover})}} is an open-source, all-terrain and waterproof stable mobile robot. The \texttt{leo\_simulator} ROS package encapsulates the robot's URDF with a wide variety of environments. Currently, the observation space of the robot is comprised of RGB frames obtained from an onboard camera and data abstracted from an IMU. The action space is comprised of 2 real numbers, the forward velocity and the angular velocity of the rover.
\item \texttt{lsd\_rover} \footnote{\texttt{(\url{https://github.com/Mars-Rover-Manipal/Active-Suspension})}} is an open-source package maintained by Mars Rover Manipal. The rover comprises of a custom 5 bar active suspension actuated by motors. The rover is equipped with a LiDAR, an IMU and a stereocamera. The observation space includes the height of the obstacle in front, the pitch angle of the base link i.e. the chassis and an approximate of the obstacle distance from the wheel centre. The action space includes 6 real numbers, 4 of which are inputs to the actuating motors and the other 2 represent the forward and angular velocity of the rover respectively.

\end{itemize}
\section{Experimentation and Results}
The aim of this experimentation is to simulate a control code in Gazebo using simple \texttt{gym} attributes provided by the developed software stack. The environment chosen is \texttt{lsd\_force\_lidar-v0}. The motive of the control code is to process the observation data provided by the \texttt{get\_observation(self)} method, feed it into a function approximator and  predict the action space data which in turn steps the agent through one timestep in that simulation. The goal of the agent is to successfully climb randomly generated obstacles of varying dimensions by actuating its motors after processing its current observation state. Recently, machine learning has been widely investigated in control problems and applied to vehicle suspension \cite{eski2009vibration} ,\cite{si2006neural}, \cite{finalkhan}, \cite{fares2020online},\cite{konoiko2019deep},
\cite{ming2020semi}
 Owing to the stochastic behaviour of both the agent and the environment, model-free reinforcement learning is chosen to formulate the unsupervised control problem. We have tested several state-of-the-art on and off-policy  model-free Reinforcement Learning algorithms like Trust Region Policy Optimization (TRPO) \cite{schulman2015trust} , Proximal Policy Optimization \cite{schulman2017proximal}, Twin Delayed Deep Deterministic Policy Gradient \cite{dankwa2019twin}, Asynchronous Advantage Actor Critic (A3C) \cite{mnih2016asynchronous} and Deep Deterministic Policy Gradient (DDPG) \cite{lillicrap2015continuous}.
As aforementioned, our goal is to develop an algorithm that enables the agent to climb randomly generated obstacles of varying depth and height. The fluctuations about the longitudinal and lateral axis of the base link termed as longitudinal and lateral stability respectively is chosen to be a suitable metric to gauge the performance of the suspension.\par
The training script is totally independent of the environment. This means that the user can change the algorithm used to learn in the training script, without having to worry about modifying the environment's structure. Thus, we see that the software developed makes it much easier for a user to simulate a robot environment on ROS using simple \texttt{gym} attributes. 

\begin{figure*}[htbp]
\begin{tabular}{cccc}
\subfloat{\includegraphics[width = 2in]{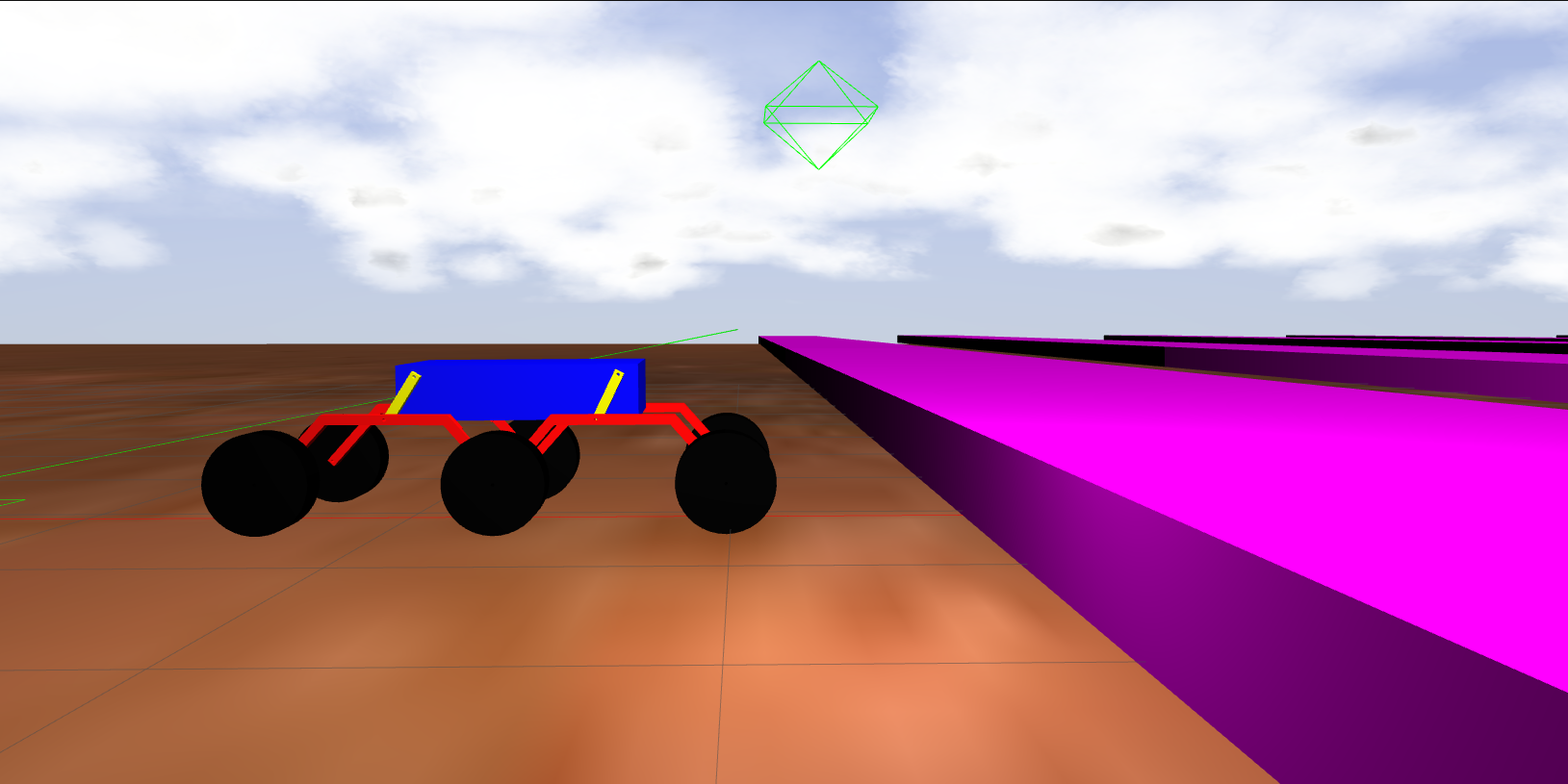}} &
\subfloat{\includegraphics[width = 2in]{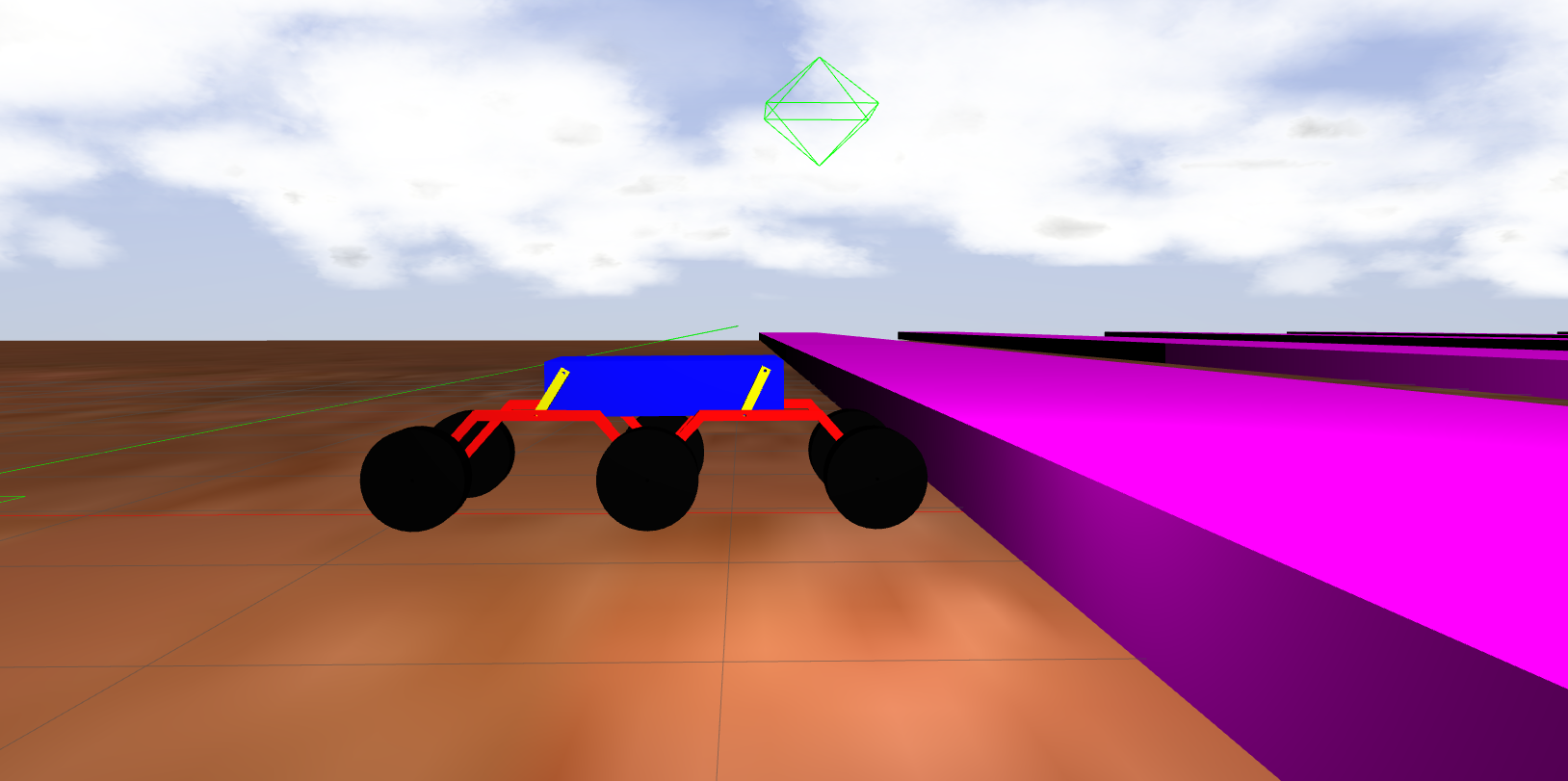}} &
\subfloat{\includegraphics[width = 2in]{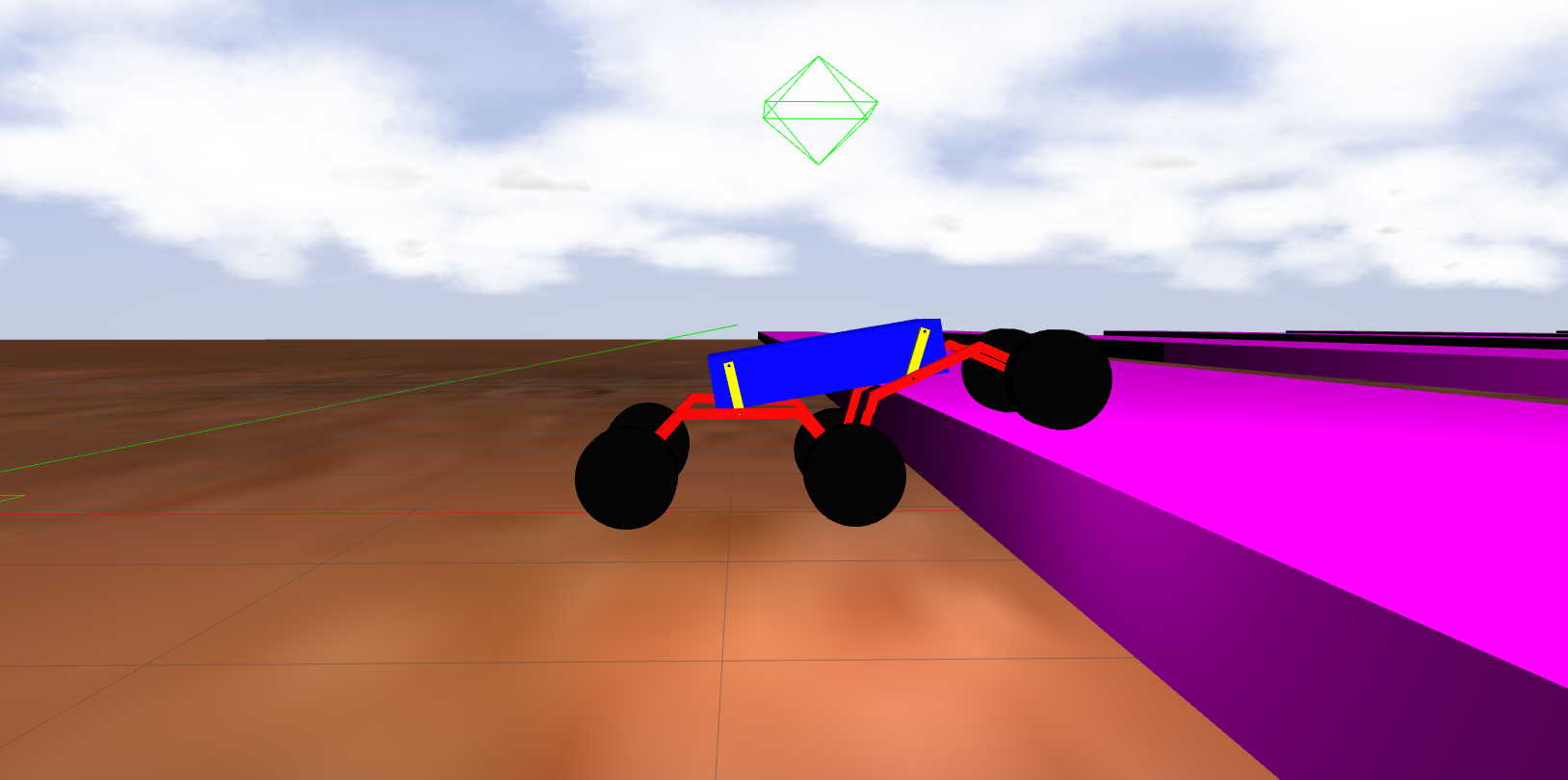}}\\
\subfloat{\includegraphics[width = 2in]{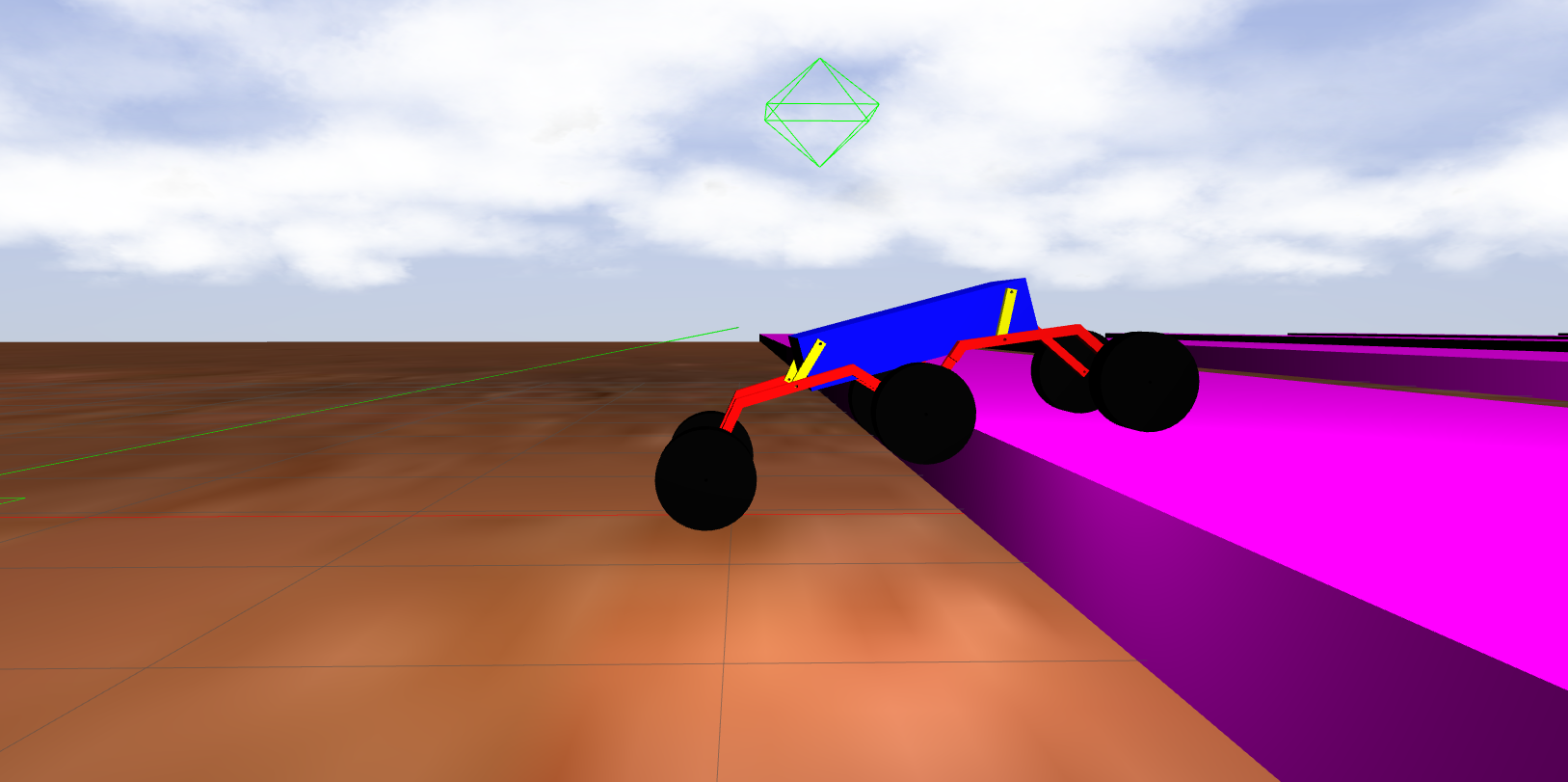}} &
\subfloat{\includegraphics[width = 2in]{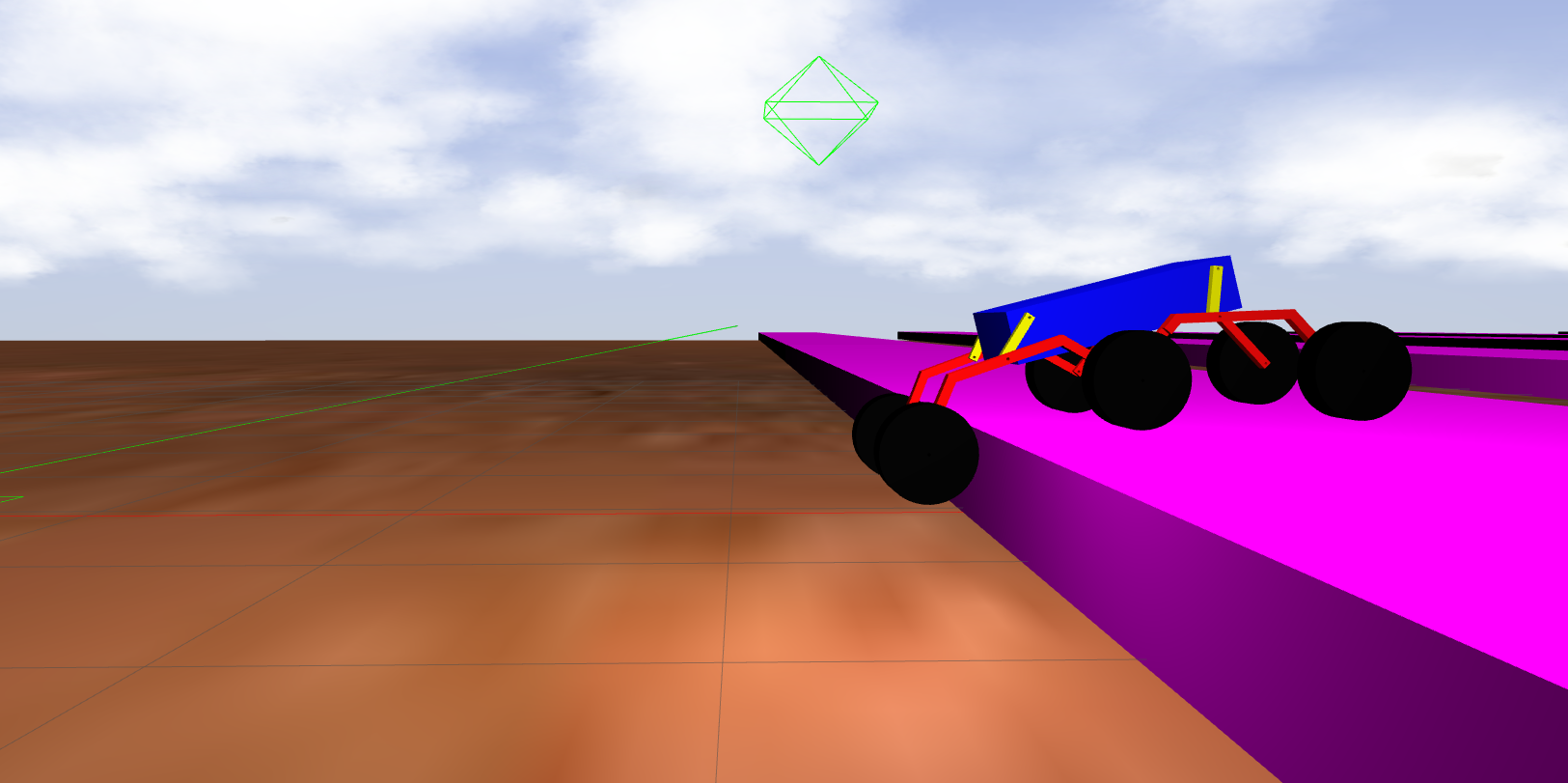}} &
\subfloat{\includegraphics[width = 2in]{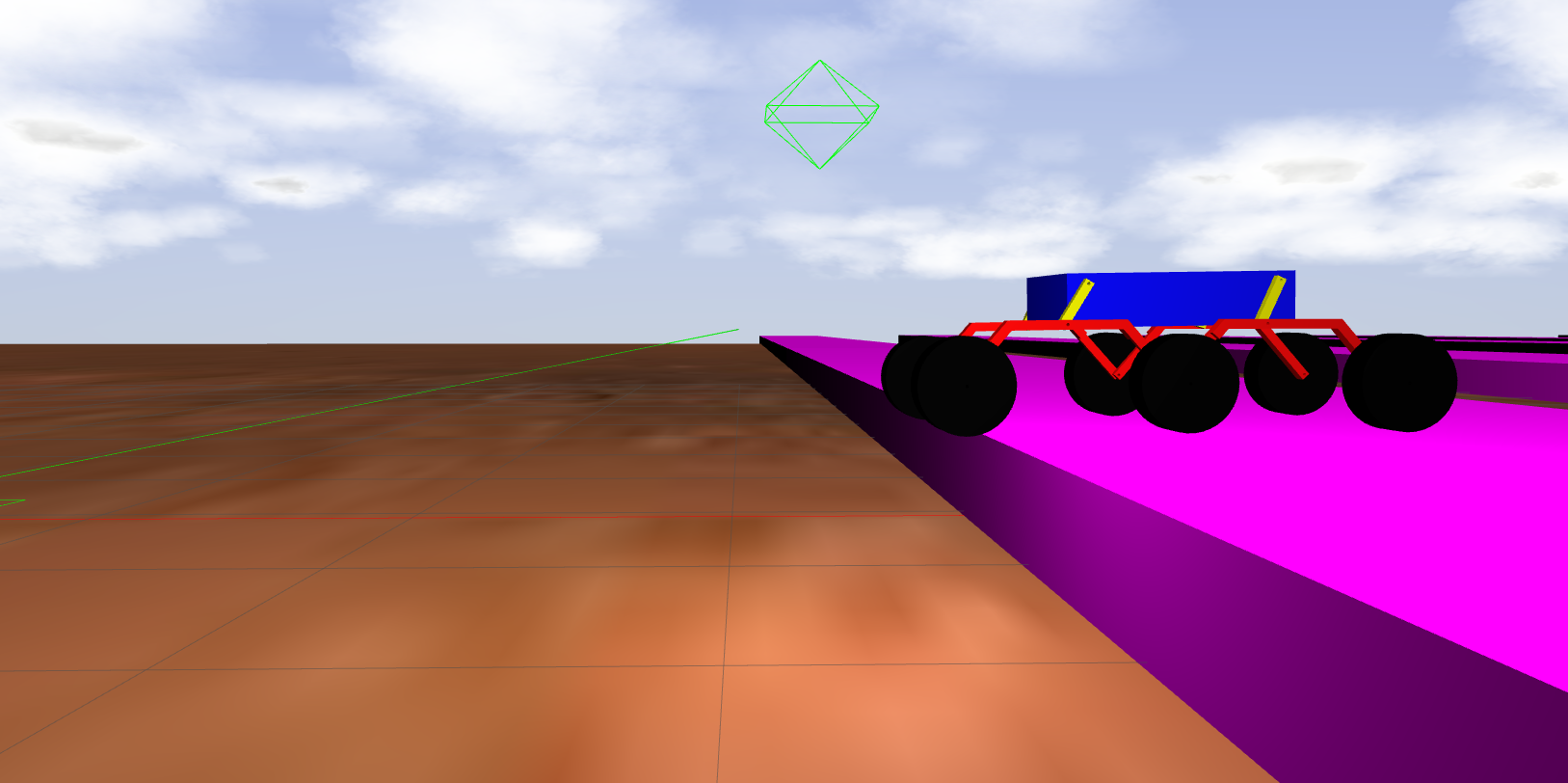}}\\
\end{tabular}
\caption{Trained model - PPO \& TD3 in action. The figure demonstrates how the trained model enables the rover to climb over a randomly generated obstacle of considerable size.}
\end{figure*}

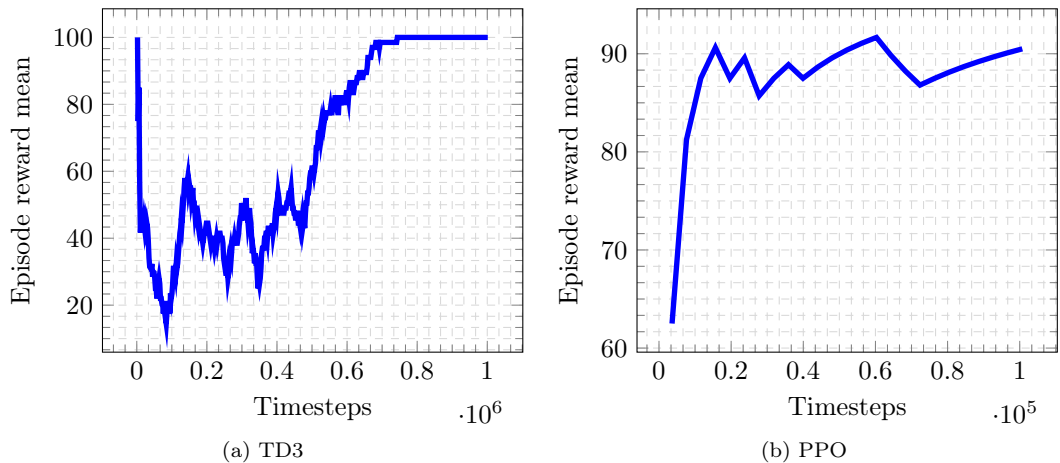
\begin{figure*}[!h]
\begin{center}
    \subfloat[TD3]{
    \begin{tikzpicture}
      \begin{axis}[
          width=2.8 in, % Scale the plot to \linewidth
          grid=both, % Display a grid
          grid style={dashed,gray!30}, % Set the style
          minor tick num=5,
          xlabel=Timesteps , % Set the labels
          ylabel= Episode reward mean]
        \addplot+ [line width=2pt,mark=none] 
        table[x= Step,y= Value, col sep=comma] {run-PPO_2-tag-rollout_ep_rew_mean.csv};
      \end{axis}
    \end{tikzpicture}
}
\hspace{0cm}
    \subfloat[PPO]{
    \begin{tikzpicture}
      \begin{axis}[
          width=2.8 in, % Scale the plot to \linewidth
          grid=both, % Display a grid
          grid style={dashed,gray!30}, % Set the style
          minor tick num=5,
          xlabel=Timesteps , % Set the labels
          ylabel= Episode reward mean]
        \addplot+ [line width=2pt,mark=none] 
        table[x= Step,y= Value, col sep=comma] {run-TD3_3-tag-rollout_ep_rew_mean.csv};
      \end{axis}
    \end{tikzpicture}
    }
\caption{Average reward plotted as a function of timesteps.}
\label{fig: trainingresults}
\end{center}
\end{figure*}
\section{Future Work}
Currently, the developed software stack is being tested to be deployed for the Virtual Mars Rover Challenge.
The future work involves us extending the developed toolkit by integrating the \texttt{gym-gazebo2} \cite{lopez2019gym} stack. \texttt{gym\_gazebo2} provides the same functionalities as \texttt{gym-gazebo} except that it facilitates integration of Gazebo with ROS 2. Furthermore, the setup script currently only generalizes to mobile ground robots. More environments integrated with open source autopilot systems like APM, PX4, ArduPilot, etc will be provided in later versions.

\section{Acknowledgements}
I would like to thank Mars Rover Manipal (\texttt{\url{http://marsrovermanipal.com/}}), for providing me with the necessary resources to successfully conduct this research. 
\bibliographystyle{plain}
\bibliography{SampleReferencesForExtendedAbstract}

\end{document}